\documentclass[aps,prl,twocolumn,floatfix,10pt,superscriptaddress]{revtex4}
\usepackage{graphicx}
\usepackage{amsmath}
\usepackage{color}
\begin{document}

\title{Depleted Depletion Drives Polymer Swelling in Poor Solvent Mixtures}

\author{Debashish Mukherji}
\email[]{mukherji@mpip-mainz.mpg.de}
\affiliation{Max-Planck Institut f\"ur Polymerforschung, Ackermannweg 10, 55128 Mainz, Germany}
\author{Carlos M. Marques}
\affiliation{Institut Charles Sadron, Universit\'e de Strasbourg, CNRS, Strasbourg, France}
\author{Torsten Stuehn}
\affiliation{Max-Planck Institut f\"ur Polymerforschung, Ackermannweg 10, 55128 Mainz, Germany}
\author{Kurt Kremer}
\email[]{kremer@mpip-mainz.mpg.de}
\affiliation{Max-Planck Institut f\"ur Polymerforschung, Ackermannweg 10, 55128 Mainz, Germany}

\date{\today}


\begin{abstract}
Macromolecular solubility in solvent mixtures often exhibit striking and paradoxical nature. 
For example, when two well miscible poor solvents for a given polymer are mixed together, the same 
polymer may swell within intermediate mixing ratios. We combine computer simulations and theoretical 
arguments to unveil the first microscopic, generic  origin of this collapse-swelling-collapse scenario. 
We show that this phenomenon naturally emerges at constant 
pressure in mixtures of purely repulsive components, especially when a delicate balance of the 
entropically driven depletion interactions is achieved.
\end{abstract}

\maketitle

It has been commonly observed that a polymer can collapse in a mixture of two competing, well miscible good solvents,
while the same polymer remains expanded in these two individual components. This phenomenon is commonly known 
as co-non-solvency \cite{wolf78macchem,schild91mac,WuPRL01,hiroki01polymer,kiritoshi03,lund04mac,mukherji13mac,mukherji14natcom,mukherji15jcp,freed15jcp}. 
However, it has also been observed that a polymer can be collapsed in two different poor solvents, 
whereas it is ``better" soluble in their mixtures \cite{masegosa84mac,hoogenboom10ajc,lee14pol,yu15acsmaclet}. 
Thus far a multitude of specific, system dependent explanations hindered the emergence of a 
clear physical picture of these two intriguing phenomena. While the phenomenon of co-non-solvency has been
recently brought onto a firmer ground of a generic explanation \cite{mukherji14natcom,mukherji15jcp}, no equivalent understanding of the 
collapse-swelling-collapse behavior has yet been achieved. In this work we propose the microscopic, generic 
picture of this collapse-swelling-collapse behavior in poor solvent mixtures.

In a standard poor solvent, starting from a good solvent condition, an increase of the effective attraction between the monomers first brings 
the polymer into $\Theta-$conditions, where the radius of gyration scales as $R_{\rm g} \sim N_l^{1/2}$ with $N_l$ 
being the chain length \cite{degennesbook,desclobook}. Upon further increase of the attraction, a polymer then collapses into 
a globular state. The resultant collapsed globule can be understood by balancing the negative second virial osmotic contributions 
and the three body repulsions. The effective attraction, between monomers of a polymer, originates
when the solvent particles repel polymer beads more than the repulsion between 
two monomers \cite{degennesbook,desclobook}. This attraction between monomer beads is referred 
to as depletion induced attraction, well known from colloidal stability \cite{lekerbook}. In this case, the resulting isolated 
polymer conformation can be well described by the Porod scaling 
law of the static structure factor $S(q) \propto q^{-4}$, presenting a compact spherical globule. 
Interestingly, even if a polymer exhibits poor solvent conditions in two different solvents, 
it can possibly be somewhat swollen by intermediate mixing ratios of the two poor solvents. A system that shows this collapse-swelling-collapse 
scenario is poly(methyl methacrylate) (PMMA) in aqueous alcohol mixtures. More specifically, water and alcohol are {\it almost} 
perfectly miscible and individually poor 
solvents for PMMA. However, PMMA shows improved solubility within the intermediate mixing concentrations of aqueous 
alcohol and/or other solvent mixtures \cite{masegosa84mac,hoogenboom10ajc,lee14pol,yu15acsmaclet}. 

In this work, we aim to (1) devise a thermodynamically consistent generic 
(chemically independent) model for a polymer in poor solvent mixtures, such that solubility of many polymers
can be explained within a simplified (universal) physical concept, (2) develop a microscopic understanding of the 
collapse-swelling-collapse scenario, and (3) investigate if a polymer in 
mixed poor solvents can really reach a fully extended good solvent chain.
To achieve the above goals, we combine generic molecular dynamics, all-atom simulations and theoretical arguments to study 
polymer solvation in poor solvent mixtures.

Our generic simulations are based on the well-known bead-spring model of polymers \cite{kremer90jcp}.
A bead-spring polymer $p$ is solvated in mixed solutions composed of two components, solvent $s$ and cosolvent $c$, respectively. 
The mole fraction of the cosolvent component $x_c$ is varied from 0 (pure $s$ component) to 1 (pure $c$ component). 
Simulations are performed using the ESPResSo++ molecular dynamics package \cite{espresso}. We have 
also performed all-atom simulations using the GROMACS molecular dynamics package \cite{gro}.
The details about generic simulations and all-atom force field parameters are given in the electronic supplementary material \cite{aux}. 

\begin{figure}[ptb]
\includegraphics[width=0.46\textwidth,angle=0]{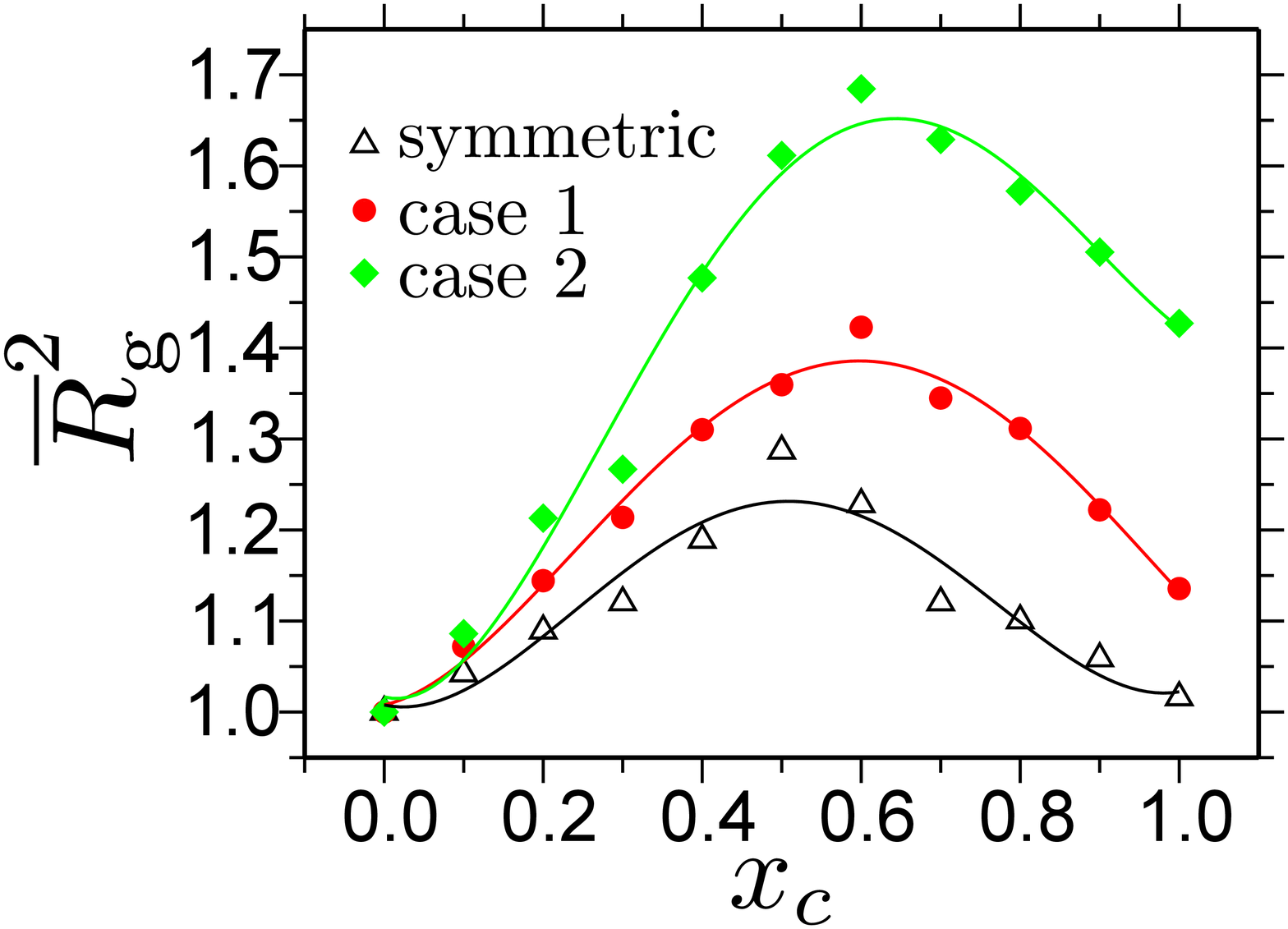}
\caption{Normalized squared radius of gyration ${\overline R}_{\rm g}^2 = \left<R_{\rm g}^2\right>/\left<R_{\rm g}(x_c = 0)^2\right>$ 
as a function of cosolvent molar concentration $x_c$. Results are shown for the generic simulations and for three different cases. 
The parameters specific details of generic cases are listed in the electronic supplementary material \cite{aux}. 
The results are shown for a chain length of $N_l = 30$. Lines are polynomial fits to the data that are drawn to guide the eye.
\label{fig:rg}}
\end{figure}

\begin{figure}[ptb]
\includegraphics[width=0.22\textwidth,angle=0]{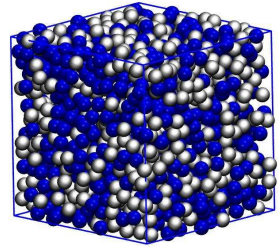}
\caption{%Part (a) presents $\rho_{\rm total}$ and pressure $p$ for the generic model as a function full cosolvent mole fraction $x_c$.
%The line is drawn based on Eq.~\ref{eq:volume}, with $\rho_{\rm total} = 1/v$.
%Parts (b) show s
Simulation snapshots of the generic system presenting bulk solution for a $x_c = 0.5$ 
mixture.
\label{fig:phase_ss}}
\end{figure}

In Fig.~\ref{fig:rg} we summarize results for the normalized squared radius of gyration 
${\overline R}_{\rm g}^2 = \left<R_{\rm g}^2\right>/\left<R_{\rm g}(x_c = 0)^2\right>$ as a function of $x_c$ from the generic model
and for three different cases described in the supplementary material \cite{aux}. 
A closer look at the symmetric case of two almost perfectly miscible solvents (black $\triangle$) show that 
$-$ while the pure solvent ($x_c = 0$) and the pure cosolvent ($x_c = 1$) are equally poor solvents for the polymer, 
the same polymer swells within the intermediate cosolvent compositions reaching maximum swelling of ${\overline R}_{\rm g}^2$ by $\sim 20\%$
at around $x_c = 0.5$. How could this be? In this context, given that this is a case of standard poor solvent collapse, 
the polymer conformations are given by depletion forces (or depletion induced attraction) \cite{lekerbook}. 
Using simple depletion arguments, when both solvent and cosolvent are 
equally repulsive, polymer conformation should remain unaltered over the full range of $x_c$. 
Here, however, when cosolvents are added into polymer-solvent system (such as the addition of alcohol in PMMA-water system),
the addition of cosolvents not only deplete monomers, but also deplete solvents. This leads to an effective double 
depletion effect that ultimately gives rise to a reduced depletion around $x_c = 0.5$ and thus is consistent with 
the swelling of the polymers. Note that the depletion forces are intimately linked to the total density of system $\rho_{\rm total}$ and, therefore,
reduced depletion forces should also be associated with the reduced $\rho_{\rm total}$.
Interestingly, when looking into aqueous-alcohol mixtures, it becomes clear that $\rho_{\rm total}$ of the solution at 
constant pressure $P$ does not change linearly with changing composition (or $x_c$) \cite{perera06jcp}. 
Instead, they exhibit a minimum in $\rho_{\rm total}$ around $x_c = 0.5$ (see Fig. 1 in electronic supplementary material \cite{aux}),
indicating that the miscibility is only {\it almost} perfect. 
In our generic simulations, we have tuned solvent-cosolvent (repulsive) interaction strength $\epsilon_{sc}$ such that the generic 
system can reproduce the weak density dip observed in the aqueous alcohol mixtures at constant $P$, see Fig.~\ref{fig:phase_ss}(a).
Furthermore, the system parameters are chosen such that the bulk solution remains deep into the miscible state far from the 
phase separation. The representative simulation snapshot is shown in Fig.~\ref{fig:phase_ss}(b) for a $50-50$ solvent-cosolvent mixture. 

\begin{figure}[ptb]
\includegraphics[width=0.40\textwidth,angle=0]{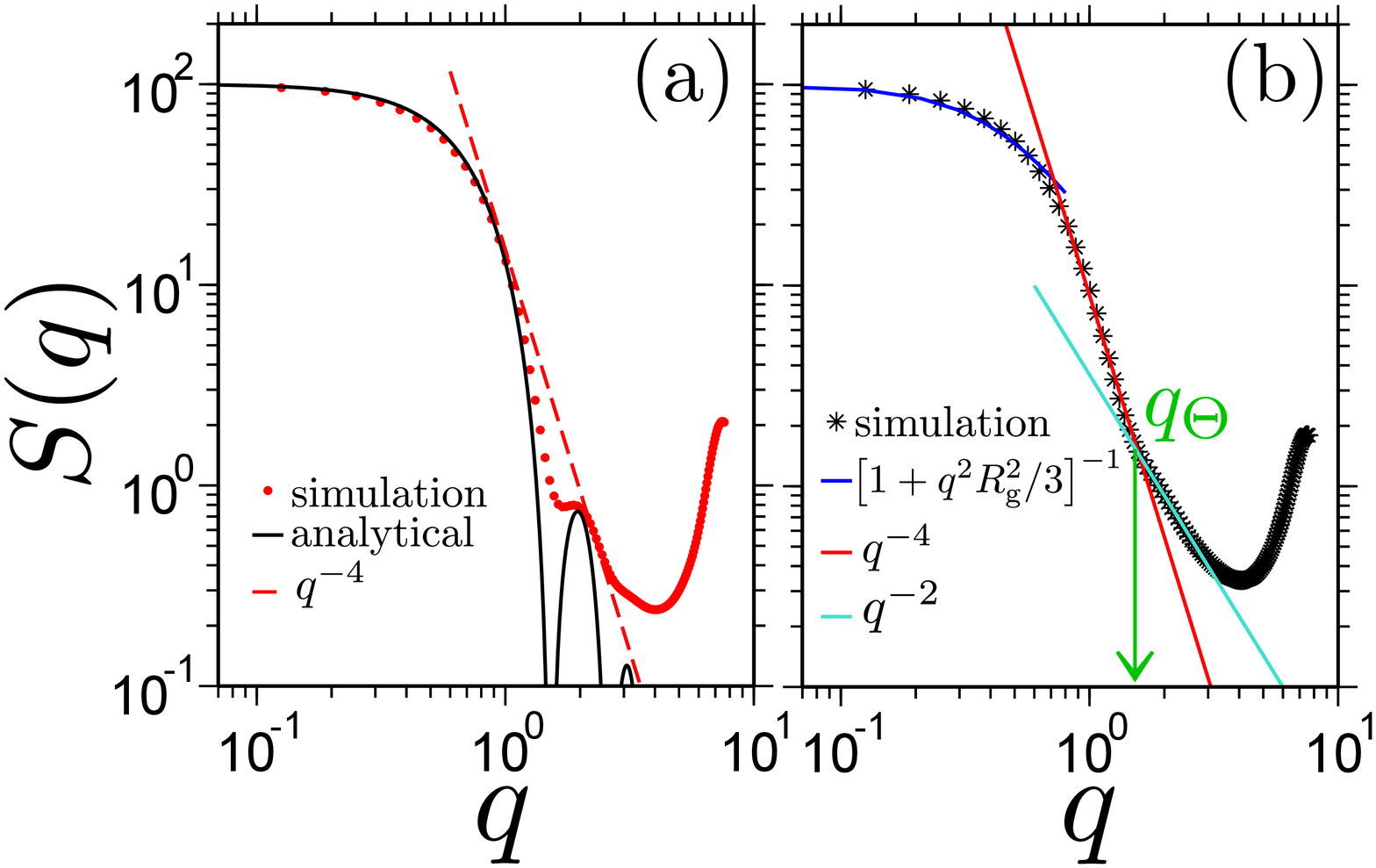}
\caption{Static structure factor $S(q)$ for a chain of length $N_l = 100$. Part (a) shows $S(q)$ at $x_c = 0.0$ and 
part (b) for $x_c = 0.7$. In part (a) we also include the analytical expression for the sphere scattering. 
In part (b) red and green lines are power law fits to the data at different length scales. Blue line represents 
Gruiner region for $q \to 0$ (for large length scales). Vertical arrow indicates the effective 
$\Theta-$blob size at $q = q_{\Theta}$, estimated using $\ell_{\Theta-{\rm blob}} = 2\pi/q_{\Theta}$.
Note to get better estimation of the cross-over scaling regime, $S(q)$ is calculated from a simulation of 
a chain length $N_l = 100$. 
\label{fig:s_of_q}}
\end{figure}

When the interaction asymmetry between polymer-cosolvent $\epsilon_{pc}$ and polymer-solvent $\epsilon_{ps}$ is increased 
(see electronic supplementary material \cite{aux}), where $\epsilon_{pc}$ for case 2 $<$ case 1 $<$ symmetric case,
not only that the degree of swelling increases, but the swelling region also shifts between $0.5 < x_c < 0.9$.
This range is found to be in excellent agreement with the experimental observation of PMMA conformation 
in aqueous alcohol mixtures \cite{hoogenboom10ajc,lee14pol}. Furthermore, our case 2 is closely resembles
PMMA in aqueous methanol mixture, where we tune parameters to mimic PMMA in aqueous methanol mixtures (see electronic supplementary material \cite{aux}). 
A closer look at Fig.~\ref{fig:rg} shows that the degree of swelling, 
within the range $0.5 < x_c < 0.9$, varies between 20-65\% (or 10-30\% in $\overline{R}_{\rm g}$) depending on the interaction assymetry. 
Considering that we are dealing with combinations of poor solvents, this is a very significant swelling. 
Moreover, analyzing the simulation it becomes apparent that the polymer does not neccesarily reach 
a fully expanded configuration. A quantity that perhaps best quantifies a polymer conformation is the static structure factor $S(q)$. 
In Fig. \ref{fig:s_of_q} we present $S(q)$ for two different values of $x_c$ for the system described by case 1. 
Part (a) shows $S(q)$ of a fully collapsed chain in pure solvent ($x_c=0$) and part (b) presents maximum polymer swelling ($x_c=0.7$).
For $x_c = 0.0$, the polymer can be well described by a scaling law known for sphere scattering (Porod scattering), 
namely $S(q) \sim q^{-4}$, suggesting a fully collapsed poor solvent conformation. 
Furthermore, the data point corresponding to $x_c = 0.7$ shows more interesting polymer conformations. 
Within the range $1.5{\sigma^{-1}} < q < 3.0{\sigma^{-1}}$ an aparent scaling $S(q) \sim q^{-2}$ is observed, which 
crosses over to $S(q) \sim q^{-4}$ for $0.7{\sigma^{-1}} < q < 1.5{\sigma^{-1}}$, suggesting that 
the polymer remains globally collapsed consisting of $\Theta-$blobs. 
The crossover point $q_{\Theta}$ gives the direct measure of the effective blob size $\ell_{\Theta-{\rm blob}} = 2\pi/q_{\Theta}$. 
The largest blobs are observed when the polymer is maximally swollen.

\begin{figure}[ptb]
\includegraphics[width=0.40\textwidth,angle=0]{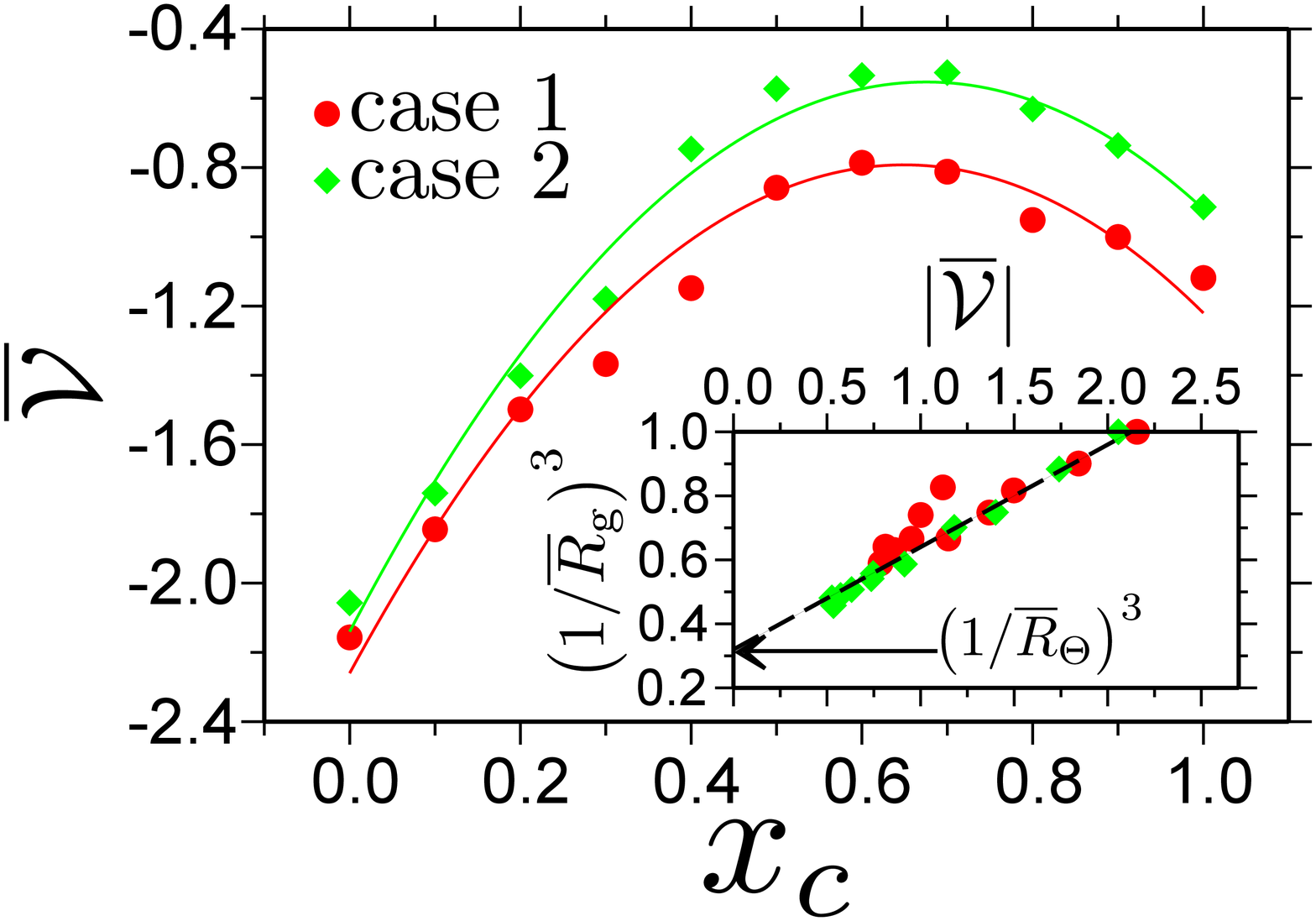}
\caption{Normalized excluded volume $\overline{\mathcal{V}} = {\mathcal{V}}/\mathcal{V}_{\rm m}$ as a function of 
cosolvent mole fraction $x_c$. Results are shown for two cases. Lines are fits to the data corresponding to Eq.~\ref{eq:fh_exvol}. 
In the inset we present $\left(1/{\overline R}_{\rm g}\right)^{3}$ as a function of normalized $\mathcal{V}$. 
Here, ${\overline R}_{\rm g} = R_{\rm g}/R_{\rm g}(x_c = 0)$ is the normalized gyration radius $R_{\rm g}$.
The line is a fit based on Eq.~\ref{eq:inter_pol}. 
\label{fig:theory}}
\end{figure}

Phase transition in polymer solutions, including the determination of changes in solvent quality leading to polymer collapse,
are conveniently described at the mean-field level by the Flory-Huggins (FH) theory and its variants.
For the case where a polymer with chain length $N_l$, at volume fraction $\phi_p$,
is dissolved in a mixture of two components $s$ and $c$, respectively,
FH theory predicts a monomer-monomer excluded volume of the form \cite{degennesbook,desclobook},
\begin{eqnarray}
\overline{\mathcal V} &=& 1 - 2 \left( 1 - x_c \right) \chi_{ps} - 2x_c \chi_{pc} + 2 x_c \left( 1 - x_c \right) \chi_{sc},
\label{eq:fh_exvol}
\end{eqnarray}
where $\chi_{ps}$ and $\chi_{pc}$ are the Flory-Huggins interaction parameters between $p-s$ and $p-c$, respectively.
The factor $\chi_{sc}$ is the parameter of $s-c$ interaction. When both solvent and cosolvent are poor solvents,
$\chi_{ps} > 1/2$ and $\chi_{pc} > 1/2$. 
In our simulations $\overline {\mathcal V} = {\mathcal V}/\mathcal{V}_{\rm m}$ is calculated using the 
expression ${\mathcal V} = 2\pi\int \left [ 1 - e^{-v(r)/k_{\rm B} T} \right] r^2 dr$.
We use $v(r) = -k_{\rm B}T\ln\left[{\rm g}(r)\right]$ as a guess of the potential of mean force (PMF), which is calculated from the radial distribution 
function between non-bonded monomers ${\rm g}(r)$. $\mathcal{V}_{\rm m} = 2.73~\sigma^3$ is the bare monomer excluded volume in the absence 
of any (co)solvent and corresponds to a monomer-monomer distance of $0.87\sigma$. 
Fitting Eq.~\ref{eq:fh_exvol} to the $\overline {\mathcal{V}}$ data in Fig.~\ref{fig:theory}, we
find $\chi_{ps} = 1.57$, $\chi_{pc} = 1.11$ and $\chi_{sc} = 1.74$ for case 1 and $\chi_{ps} = 1.62$, $\chi_{pc} = 0.95$ 
and $\chi_{sc} = 1.74$ for case 2. Consistently $\chi_{sc}$ values for both cases are similar, since this parameter is 
independent of polymer solvent interactions. Note also that standard FH predictions assume that solvent and 
cosolvent $-$ and also polymer $-$ are mixed at constant volume, whereas our simulations and experiments are performed at a constant 
pressure. In the following we derive a FH expression for the $\mathcal{V}$ values at constant $p$, which
predicts reduced effective values for $\chi_{sc}$ dependent on $p$.

In our simulations, we only consider polymer under infinite dilution $\phi_{p} \to 0$ and the large majority of the volume 
is occupied by solvent-cosolvent mixture. Therefore, we concentrate our analysis on the binary mixture. 
Additionally, we also consider that the pure reference solvent and cosolvent systems are identical, but 
that $s-c$ interactions are distinct from those for $s-s$ and $c-c$. For this case, the total free energy is written as
\begin{eqnarray}
\frac {{\mathcal F}   {v}}{\kappa_{\rm B} T}
&=&\frac{{v} {\mathcal F}_s ({v})}{\kappa_{\rm B} T}  +  x_c  \ln (x_c)+ ( 1 - x_c ) \ln(1 - x_c) \nonumber\\
&+& \chi_{sc}({v}) x_c (1 -x_c),
\label{eq:fsctotal}
\end{eqnarray}
where ${\mathcal F}_s ({v})$ is the volume-dependent free-energy of the pure solvent or cosolvent systems
and where we consider the explicit dependence of $\chi_{sc}$ on system volume.
Note that since experiments and simulations are performed at constant number of molecules ${\mathcal N}$,
the total volume of the system $V$ is simply given by $V = {\mathcal N} {v}$.
For a given external pressure $P$, the molar volume ${v}$ is thus controlled by,
$P=P_s({v}) - {\kappa_{\rm B} T} x_c (1 -x_c) {\partial\chi_{sc}({v})}/{\partial{v}}$
with $P_s({v}) = - \partial {v} {\mathcal F}_s / \partial  {v}$ being the pressure of
the reference system. If one assumes a small variation of the molar volume of the solvent-cosolvent mixture with respect to that of the reference system, one gets
\begin{eqnarray}
{v}= {v_{\rm o}} \left[1 + \zeta \  x_c (1 -x_c)\right],
\label{eq:volume}
\end{eqnarray}
where $\zeta = {\kappa_{\rm B} T}/{v}\ \partial\chi_{sc}({v})/
\partial{v} \left[ \partial P_s({v}) /\partial {v} \right]^{-1}$
measures the relative sensitivity of the interaction parameter and reference pressure to ${v}$.
In supplementary material \cite{aux} we show $P_{s}$ as a function of $v$ that gives an estimate 
of $\partial P_s({v}) /\partial {v} = 20\epsilon/\sigma^6$.
Eq. \ref{eq:volume} describes well the observed density variation of the generic model in Fig. \ref{fig:phase_ss}(a)
with $\zeta=0.26$. Note that $\rho_{\rm total}$ and molar volume $v$ are simply related
by $\rho_{\rm total} = 1/{v}$. Also to first order in $({v} - {v_{\rm o}})/{v_{\rm o}}$, which for our generic model is of the order of 10\%, one gets
\begin{eqnarray}
\chi_{sc}({v})&=& \chi_{sc}({v_{\rm o}})
+{v} \frac{\partial\chi_{sc}({v})}{\partial{v}}{\bigg{|}_{x_c \to 0}} \zeta x_c (1-x_c),
\label{eq:chi}
\end{eqnarray}
thus showing that the interaction parameter is only perturbed to second order in beta. This will lead to an
effective expression $\chi_{sc}({v})= \chi_{sc}({v_{\rm o}}) - 0.089 x_c (1-x_c)$. 
Incorporating Eq.~\ref{eq:chi} into Eq.~\ref{eq:fh_exvol} gives rise to the collapse-swelling-collapse 
scenario shown in Fig. \ref{fig:theory}. Furthermore, if we choose $x_c = 0.5$, the above equation will 
lead to a $\sim2\%$ variation in $\chi_{sc}$ values with respect to the standard values calculated in NVT ensemble. 
This suggests that $-$ while the phenomenon naturally emerges at constant $P$, the constant volume (lattice models), 
though not appropriate, could still reasonably work because of a small deviation of $\chi_{sc}$ between two ensembles.
However, one should also note that on a lattice it is difficult to speak of constant $P$.

Our numerical predictions successfully account for polymer swelling 
in solutions of poor solvent mixtures. The simulations quantitatively confirm that $-$ while this fascinating polymer behavior 
is driven by purely repulsive interactions, they also reveal the subtle balance of local depletion forces and the 
bulk solution properties behind the paradoxical nature of such phenomena. 
Indeed, polymer collapse in repulsive solvents can be understood by depletion induced attractions \cite{lekerbook}. 
The dominant contribution to the depletion attraction potentials originates from the direct monomer-solvent repulsions, and 
is thus proportional to solvent number density $\rho_{\rm totat}$ dictating number of depletants. When a few
solvent molecules are replaced by cosolvents, for example a water by an alcohol, preserves, at first order in solvent density, 
a linear composition rule for the total strength of the attractions, especially for simulations under NVT ensemble.
Under these conditions one should smoothly interpolate between two polymer collapsed states, without any swelling at intermediate
compositions. Here, however, interactions between solvent components play a delicate role in dictating the depletion forces, bringing 
in contributions proportional to the square (or to even higher powers) of $\rho_{\rm total}$, see Fig. \ref{fig:phase_ss}(a).
The dominant effect of these solvent-cosolvent (depletant-depletant) interactions is to reduce the total depletion attractions.
Such effects can be understood by noticing that any (co)solvent molecule present between two monomers already depletes 
a significant number of particles from its vicinity and thus reduces the number of depletants contributing to the total 
monomer-monomer attraction. In our case reduction of polymer collapse is obtained, enhanced and tuned by a solvent-cosolvent excluded volume 
that is slightly stronger than the corresponding values for solvent-solvent and cosolvent-cosolvent molecules. 
At intermediate compositions, where solvent-cosolvent interactions are dominant, 
there is a stronger reduction of the attractive depletion forces leading to polymer collapse, thus allowing for a 
significant polymer swelling. In practice, therefore, a broad variety of polymer/solvent systems are 
expected to display such behavior provided that both repulsive solvents have mutual interactions, leading to
depletion of depletion forces.

A standard measure of the attractive forces leading to polymer collapse is provided by the monomer excluded volume $\mathcal{V}$. For
poor solvents $\mathcal{V}$ is negative, and the dimensions of the chain can be understood by balancing the second (negative) virial osmotic 
contributions and the three body repulsion \cite{degennesbook,desclobook}, leading to
\begin{equation}
\frac {{\overline{R}_{\Theta}}^{3}} {{\overline{R}_{\rm g}}^{3}} - {1} = |\overline {\mathcal{V}}|.
\label{eq:inter_pol}
\end{equation}
In the inset of Fig. \ref{fig:theory} we show $\left(1/{\overline R}_{\rm g}\right)^{3}$ as a function of $\overline {\mathcal{V}}$, where the $R_{\rm g}$ 
is taken from Fig. \ref{fig:rg}(a) and $\overline {\mathcal{V}}$ is given by the values in the main panel of Fig. \ref{fig:theory}. 
The data is well described by the theoretical prediction in Eq. \ref{eq:inter_pol}. Extrapolating the data to 
$\mathcal{V} = 0$, we estimate ${\overline R}_{\Theta} = 1.46$ (or $R_{\Theta} = 2.34\sigma$), further suggesting that the polymer remains below 
$\Theta-$conformation, even when it swells within intermediate mixing ratios. 

This collapse-swelling-collapse scenario of PMMA in aqueous alcohol appears as the opposite 
effect to that of coil-globule-coil scenario of PNIPAm in aqueous alcohol, often referred to as co-non-solvency \cite{schild91mac,WuPRL01,mukherji14natcom}. 
However, the coil-globule-coil transitions are dictated by effects of solvent and cosolvent preferential adsorption \cite{mukherji14natcom}, 
while the collapse-swelling-collapse behavior, studied here, is due to a subtle balance of depletion forces. 
Furthermore, our analysis also suggests that, contrary to the co-non-solvency effect that cannot be described by a Flory-Huggins mean-field picture, 
the mean-field behavior drives the collapse-swelling-collapse sequence in poor solvent mixtures. Here, the solvent-cosolvent interaction 
parameter $\chi_{sc}$, though quite small, plays a key role. 
Our results clarify that although collapse-swelling-collapse and co-non-solvency appear as 
two symmetric manifestations of polymer solubility, they are in fact driven by markedly different physical mechanisms.

In conclusion, we have performed molecular dynamics simulations to unveil the microscopic origin of 
polymer swelling in poor solvent mixtures. We propose a unified generic picture of the polymer 
collapse-swelling-collapse transition. This conformational transition is due to a delicate balance 
between the depletion forces and the bulk solution density at constant pressure. 
Combining Flory-Huggins type mean field picture with molecular dynamics simulations, we show that the 
polymer swelling in poor solvents is dictated by reduced depletion forces that ordinate because the 
bulk solution properties. More interestingly, these results show striking quantitative agreement with the 
experimental data obtained from PMMA solvated in aqueous alcohol and all-atom simulations. While the polymer 
swell significantly, the mostly swollen polymer structure still remains below $\Theta-$conformation. 

{\it Acknowledgment}: D.M. thanks Burkhard D\"unweg and Vagelis Harmandaris 
for many stimulating discussions, Tiago Oliveira for the help 
to build the all-atom PMMA force field and Bj\"orn Baumeier for suggesting Ref. [4] in 
supplementary material \cite{aux}. C.M.M. acknowledges Max-Planck Institut f\"ur Polymerforschung for hospitality where this work was 
initiated and performed. We thank Nancy Carolina Forero-Martinez and Hsiao-Ping Hsu for critical reading of the manuscript.


\begin{thebibliography}{22}

\bibitem{wolf78macchem}
B. A. Wolf and M. M. Willms,
Makromol. Chem. {\bf 179}, 2265 (1978).

\bibitem{schild91mac}
H. G. Schild, M. Muthukumar, and D. A. Tirrell,
Macromolecules {\bf 24}, 948 (1991).

\bibitem{WuPRL01}
G. Zhang and C. Wu,
Phys. Rev. Lett. {\bf 86}, 822 (2001).

\bibitem{hiroki01polymer}
A. Hiroki, Y. Maekawa, M. Yoshida, K. Kubota, and R. Katakai,
Polymer {\bf 42}, 1863 (2001).

\bibitem{kiritoshi03}
Y. Kiritoshi and K. Ishihara,
Sci. Technol. Adv. Mater. {\bf 4}, 93 (2003).

\bibitem{lund04mac}
R. Lund, L. Willner, J. Stellbrink, A. Radulescu, and D. Richter,
Macromolecules {\bf 37}, 9984 (2004).

\bibitem{mukherji13mac}
D. Mukherji and K. Kremer,
Macromolecules {\bf 46}, 9158 (2013).

\bibitem{mukherji14natcom}
D. Mukherji, C. M. Marques, and K. Kremer,
Nat. Commun. {\bf 5} 4882 (2014).

\bibitem{mukherji15jcp}
D. Mukherji, C. M. Marques, T. Stuehn, and K. Kremer,
J. Chem. Phys. {\bf 142}, 114903 (2015).

\bibitem{freed15jcp}
J. Dudowicz, K. F. Freed, and J. F. Douglas,
{\it J. Chem. Phys.} {\bf 143}, 131101 (2015).

\bibitem{masegosa84mac}
R. M. Masegosa, M. G. Prolongo, I. Hernandez-Feures, and A. Horta,
Macromolecules, {\bf 17}, 1181 (1984).

\bibitem{hoogenboom10ajc}
R. Hoogenboom, C. Remzi Becer, C. Guerrero-Sanchez, S. Hoeppener, and U. S. Schubert,
Aust. J. Chem. {\bf 63} 1173 (2010).

\bibitem{lee14pol}
S. M. Lee and Y. C. Bae,
Polymer {\bf 55} 4684 (2014).

\bibitem{yu15acsmaclet}
Y. Yu, B. D. Kieviet, E. Kutnyanszky, G. J. Vancso, and S. de Beer,
ACS Macro Letters {\bf 4} 75 (2015).

\bibitem{degennesbook}
P.-G. de Gennes,
{\it Scaling Concepts in Polymer Physics}
(Cornell University Press, London, 1979).

\bibitem{desclobook}
J. Des Cloizeaux and G. Jannink,
{\it Polymers in Solution: Their Modelling and Structure}
(Clarendon Press, Oxford, 1990).

\bibitem{lekerbook}
H. N. W. Lekkerkerker and R. Tuinier,
{\it Colloids and the Depletion Interaction}
(Clarendon Press, Oxford, 1990).

\bibitem{kremer90jcp}
K. Kremer and G. S. Grest,
J. Chem. Phys. {\bf 92}, 5057 (1990).

\bibitem{espresso}
J. D. Halverson, T. Brandes, O. Lenz, A. Arnold, S. Bevc, V. Starchenko, K. Kremer, T. Stuehn, D. Reith,
Comp. Phys. Comm. {\bf 184}, 1129 (2013)

\bibitem{gro}
S. Pronk, S. Pall, R. Schulz, P. Larsson, P. Bjelkmar, R. Apostolov, M. R. Shirts, J. C. Smith, P. M. Kasson, D. van der Spoel, B. Hess, and E. Lindahl,
Bioinformatics {\bf 29}, 845 (2013).

\bibitem{aux}
Electronic auxilliary material. Number to be filled by the editor. 

\bibitem{perera06jcp}
A. Perera, F. Sokolic, L. Almasy, and Y. Koga,
J. Chem. Phys. {\bf 124}, 124515 (2006).

\end{thebibliography}
\end{document}